\documentclass[twocolumn,showpacs,showkeys,amsmath,amssymb]{revtex4}

\usepackage{graphicx}
\usepackage{dcolumn}
\usepackage{bm}

\begin{document}

\title{Electronic structure of the zigzag spin-chain compound In$_2$VO$_5$}

\author{U.~Schwingenschl\"ogl}
\affiliation{Institut f\"ur Physik, Universit\"at Augsburg, 86135 Augsburg,
Germany}

\date{\today}

\begin{abstract}
Band structure calculations within the local spin-density approximation
are presented in order to investigate the electronic and magnetic
properties of the zigzag spin-chain compound In$_2$VO$_5$. The
essential structural feature of the system is a double chain of
VO$_6$-octahedra, which leads to competing intrachain and interchain
magnetic couplings. Frustration of the spin-chains is expected for
the proposed antiferromagnetic ordering at low temperatures. However,
the band calculations indicate that the experimental room temperature
crystal structure is incompatible with antiferromagnetism. Both
the intrachain and interchain coupling is found to be ferromagnetic.
\end{abstract}

\pacs{71.20.-b, 71.20.Be}
\keywords{density functional theory, band structure, magnetism, spin-chain}

\maketitle

Magnetism in low-dimensional quantum spin systems results
in fascinating physical properties when the spin ordering is frustrated
due to geometrical restrictions. In this context, the $S=1/2$ antiferromagnetic
zigzag spin-chain gives rise to one of the most fundamental model
systems for analyzing the interplay of frustration and magnetism.
Recently, In$_2$VO$_5$ has been put forward for consideration as a
promising candidate for a frustrated zigzag spin-chain. Vanadium
oxides in general are very susceptible to electronic ordering
phenomena, see \cite{goodenough71,bruckner83,us04}, for instance, and
the references given there. To be more specific,
the essential structural feature of In$_2$VO$_5$ is a double chain of
corner sharing VO$_6$-octahedra along the crystallographical $b$-axis.
As indium realizes the oxidation state In$^{3+}$, the vanadium
ions are left with a single electron in the $3d$ shell. This formal
V$^{4+}$ valence with a 3$d^1$ electronic configuration comes along
with $S=1/2$ spins at the vanadium sites, separated by non-magnetic
oxygen sites. Because of nearest and next-nearest neighbour V-V
interactions of the same order of magnitude, competence between the
intrachain and interchain exchange coupling is typical for In$_2$VO$_5$.

\begin{figure}
\includegraphics[width=0.4\textwidth]{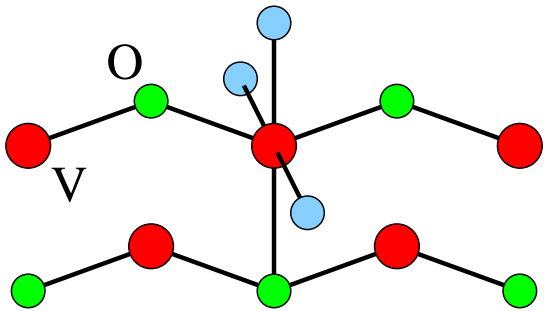}
\caption{(Color online) Schematic structure of the zigzag VO-chains
in In$_2$VO$_5$. The magnetic V sites are surrounded by distorted
O-octahedra, sharing corners along the chains.}
\label{fig1}
\end{figure}
The compound crystallizes in the simple orthorhombic space group
$Pnma$, where a unit cell comprises four formula units. Senegas
{\it et al.} \cite{senegas75} have obtained the lattice parameters
$a=7.232$\,\AA, $b=3.468$\,\AA, and $c=14.82$\,\AA\ by means of single
crystal x-ray analysis at room temperature. Figure \ref{fig1}
illustrates the spacial arrangement of the zigzag
VO-chains. The coordination polyhedron of the magnetic V sites is a
distorted O-octahedron with one strongly elongated VO-bond of length
2.23\,\AA. This bond connects two adjacent VO-chains, therefore giving
rise to the characteristical double chain geometry. The VO-bond in
trans-position accordingly is shortened to 1.76\,\AA. Moreover, the
intrachain bond length amounts to 1.81\,\AA\ in both directions and
the remaining VO-bonds in the equatorial plane of the coordination
polyhedron have lengths of 2.01\,\AA\ and 2.03\,\AA. All V sites
are crystallographically equivalent, as are the intrachain O sites.
Within the zigzag VO-chains we have V-O-V bond angles of 146$^\circ$,
thus considerable deviations from a straight line configuration.
In contrast, the V-O-V bond angles between V sites in adjacent
chains amount to only 107$^\circ$. Nearest neighbour magnetic sites
therefore are located in different VO-chains, whereas intrachain
V sites are next-nearest neighbours.

Sign and strength of the magnetic coupling constants in In$_2$VO$_5$
very recently have been analyzed by Volkova \cite{volkova07} based on
a phenomenological theoretical method for quantitatively estimating
the magnetic coupling in low-dimensional
crystalline compounds. The only input into the calculation is the
crystal structure \cite{volkova05}. This crystal chemical approach results in
an antiferromagnetic nearest neighbour interaction given by the coupling
constant $J_1=-0.9$\,mRyd. Competing magnetic interaction along the zigzag
VO-chains is likewise antiferromagnetic, with coupling constant
$J_2=-1.6$\,mRyd. Despite two paths for the nearest
neighbour exchange and a single path for the intrachain exchange, see
figure \ref{fig1}, the next-nearest neighbour magnetic coupling therefore
is found to dominate. Furthermore, since the ratio $J_2/J_1=1.68$
exceeds the Majumdar-Ghosh point 0.5, the zigzag spin-chains in
In$_2$VO$_5$ are expected to be frustrated.

The following electronic structure results for In$_2$VO$_5$ rely on
the scalar-relativistic augmented spherical wave (ASW) method
\cite{eyert07}. The implementation in use particularly accounts for
the non-spherical contributions to the charge density inside the atomic
spheres. The structural input for the calculation is taken from
Senegas {\it et al.} \cite{senegas75}. For a correct representation
of the crystal potential in voids of the In$_2$VO$_5$ structure the
physical spheres have to be complemented by additional augmentation
spheres at carefully selected interstitial sites. It turns out that
it is sufficient to dispose 84 additional spheres from 15
crystallographically inequivalent classes in order to keep the linear overlap of the physical
spheres below 15\% and the overlap of any pair of spheres below 20\%.
Since we have 32 physical spheres, the unit cell entering the
calculation thus comprises 116 augmentation spheres in total. The
basis set taken into account in the secular matrix consists of In
$5s$, $5p$, $4d$, V $4s$, $4p$, $3d$, and O $2s$, $2p$ states,
as well as states of the additional augmentation spheres. During the
course of the band structure calculation the Brillouin zone is sampled with an
increasing number of up to 56 {\bf k}-points in the irreducible
wedge, which ensures convergence of the results with respect to the
fineness of the {\bf k}-space grid. For the exchange-correlation
functional the Vosko-Wilk-Nusair parametrization is used. As long as
the augmentation spheres are selected carefully and
the crystal structure is not altered, the ASW method is
highly reliable for the comparison of magnetic energies.

\begin{figure}
\includegraphics[width=0.45\textwidth,clip]{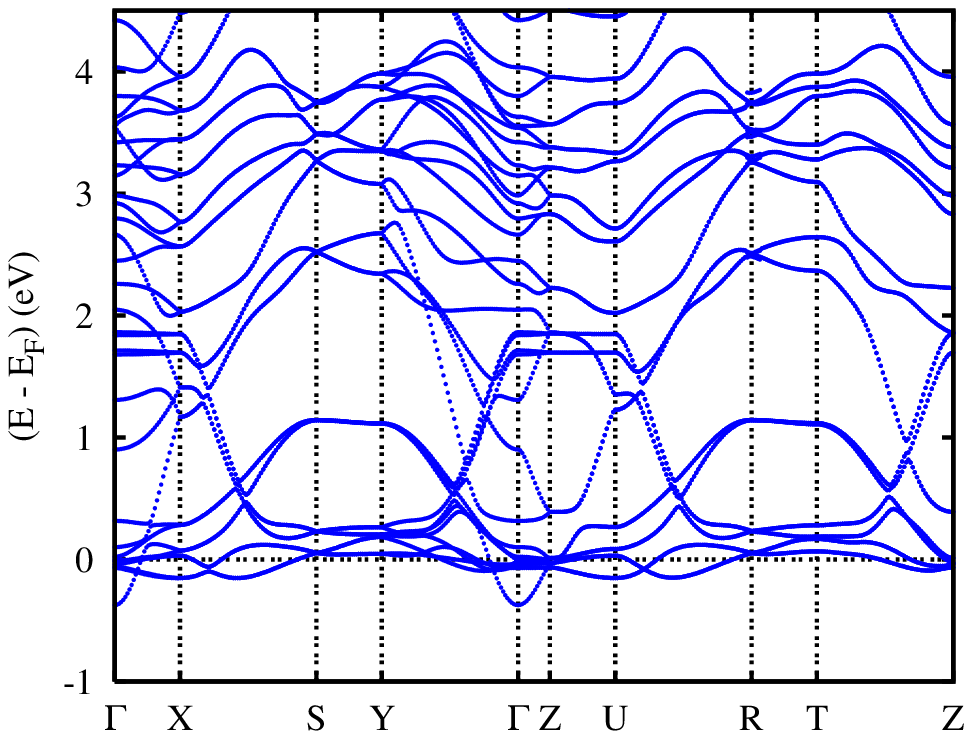}\\
\includegraphics[width=0.45\textwidth,clip]{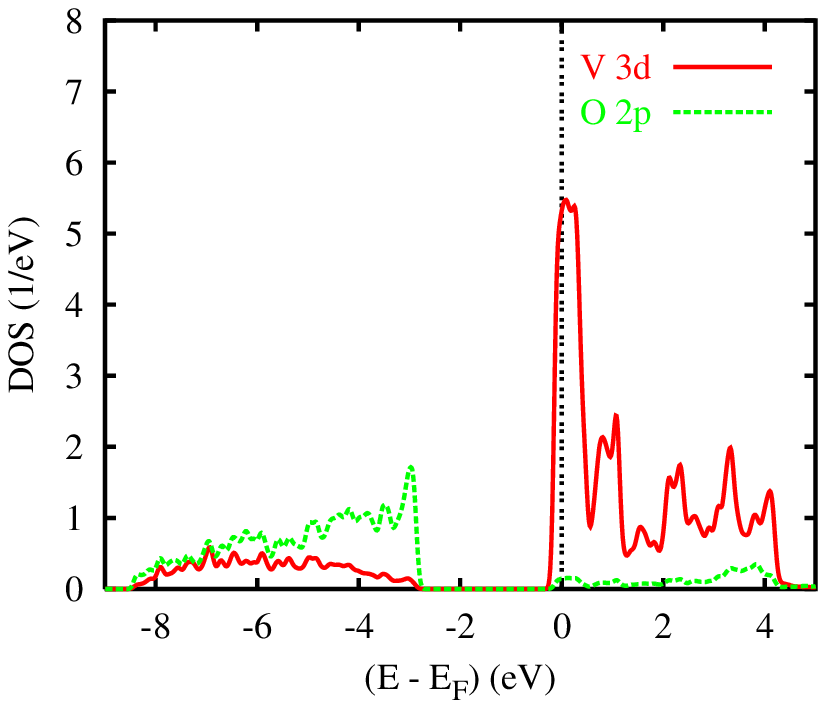}
\caption{(Color online) Band structure and partial V $3d$
and O $2p$ DOS (per atom), as resulting from a spin-degenerate
calculation.}
\label{fig2}
\end{figure}
We start the discussion of the electronic structure of In$_2$VO$_5$
by addressing results of a spin-degenerate band structure calculation.
They allow us to study general issues concerning the anisotropy of the
electronic states, the chemical bonding, and hybridization effects.
Afterwards we will investigate the magnetic coupling by spin-polarized
calculations for various spin patterns. Figure \ref{fig2} shows the
band structure for the spin-degenerate case along selected
high symmetry lines in the first Brillouin zone of the simple
orthorhombic lattice, where the high symmetry points are defined by 
the standard reciprocal lattice vectors $\Gamma=(0,0,0)$, $X=(\frac{1}{2},0,0)$,
$S=(\frac{1}{2},\frac{1}{2},0)$, $Y=(0,\frac{1}{2},0)$, $Z=(0,0,\frac{1}{2})$,
$U=(\frac{1}{2},0,\frac{1}{2})$, $R=(\frac{1}{2},\frac{1}{2},\frac{1}{2})$,
and $T=(0,\frac{1}{2},\frac{1}{2})$. In the vicinity of the Fermi
energy, we have several electronic bands revealing little dispersion
throughout the first Brillouin zone. Because these bands originate
from the V $3d$ states, they are responsible for a remarkable peak
in the V $3d$ density of states (DOS), compare the DOS curves in
figure \ref{fig2}. In the energy range shown, we have almost only
contributions from the V $3d$ and O $2p$ states. Fully occupied
In $4d$ bands give rise to a pronounced structure around $-15$\,eV,
with respect to the Fermi level.

The gross features of the partial V $3d$ and O $2p$ DOS are typical
for compounds based on VO$_6$-octahedra. As to be expected from a
molecular orbital picture, we can identify two groups of bands in the
energy ranges from $-8.5$\,eV to $-2.8$\,eV and from $-0.2$\,eV to
4.2\,eV. Interaction between V $3d$ and O $2p$ atomic orbitals leads
to bonding and antibonding molecular states. The bonding bands are
fully occupied, whereas the antibonding bands cross the Fermi level
and cause In$_2$VO$_5$ to be a metal at room temperature. Even though
the bonding and antibonding states are dominated by oxygen and
vanadium contributions, respectively, non-vanishing admixtures of the
other states are present in figure \ref{fig2}. They trace back to
significant VO-hybridization, particularly between orbitals mediating
$\sigma$-type overlap.

\begin{table}
\begin{tabular}{c|c|c}
intrachain coupling & interchain coupling & energy gain\\\hline
&&\\[-0.2cm]
fe & fe & 12\,mRyd\\
fe & af & 10\,mRyd\\
af & fe & 4\,mRyd
\end{tabular}
\vspace{0.5cm}
\caption{Comparison of the energy gain (per V site)
due to the exchange coupling for various spin patterns.}
\label{tab1}
\end{table}

For investigating the magnetic coupling in In$_2$VO$_5$, we have to
consider the following three spin patterns, since nearest and
next-nearest neighbour exchange interaction is relevant. First, we
assume the magnetic coupling to be ferromagnetic both along the
VO-chains and between neighbouring chains. Afterwards, we assume
either the intrachain or the interchain coupling to be
antiferromagnetic, while keeping the other coupling ferromagnetic.
In each case, spin-polarized band structure calculations result in
a lowering of the total energy as compared to the spin-degenerate
solution. Values for the energy gain per magnetic site are summarized
in table \ref{tab1}. The largest energy gain of 12\,mRyd is obtained
when both the nearest and next-nearest neighbour exchange interaction
is ferromagnetic. With respect to this value, intrachain and interchain
antiferromagnetic coupling raises the energy by 2\,mRyd and 8\,mRyd,
respectively. The magnetic ground state of In$_2$VO$_5$ hence is
found to be ferromagnetic, i.e.\ the room temperature crystal structure
of Senegas {\it et al.} \cite{senegas75} is incompatible with
antiferromagnetism within the VO-chains as well as between the chains.

\begin{figure}
\includegraphics[width=0.45\textwidth,clip]{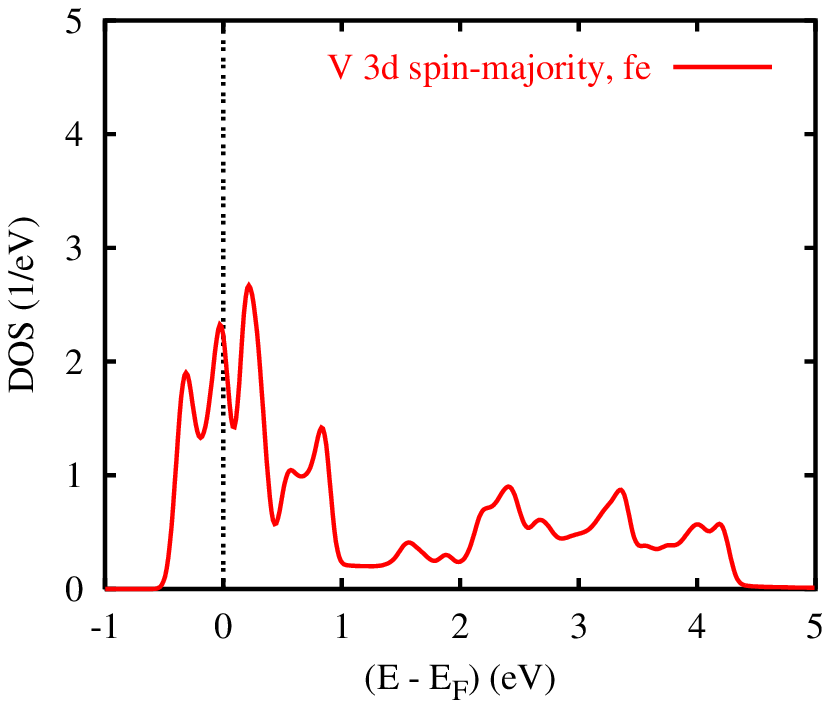}\\
\includegraphics[width=0.45\textwidth,clip]{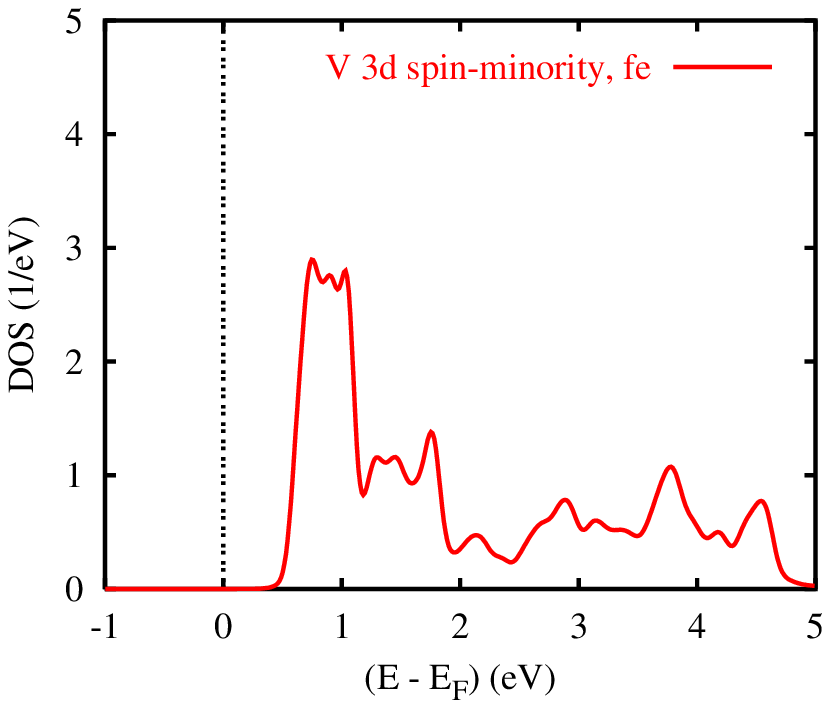}
\caption{(Color online) Partial V $3d$ spin-majority (top) and
spin-minority (bottom) DOS (per atom) for ferromagnetic
intrachain coupling.}
\label{fig3}
\end{figure}
\begin{figure}
\includegraphics[width=0.45\textwidth,clip]{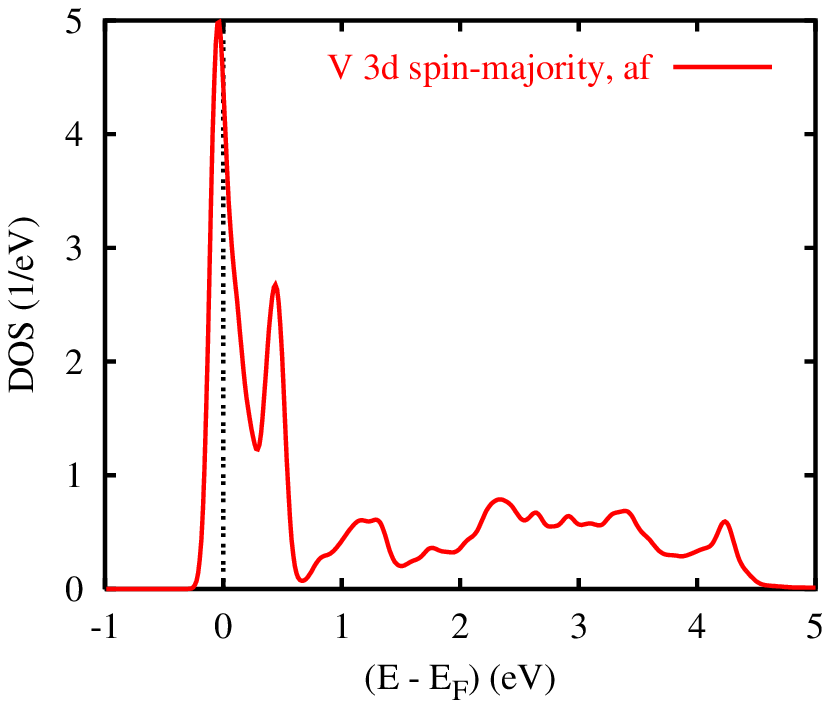}\\
\includegraphics[width=0.45\textwidth,clip]{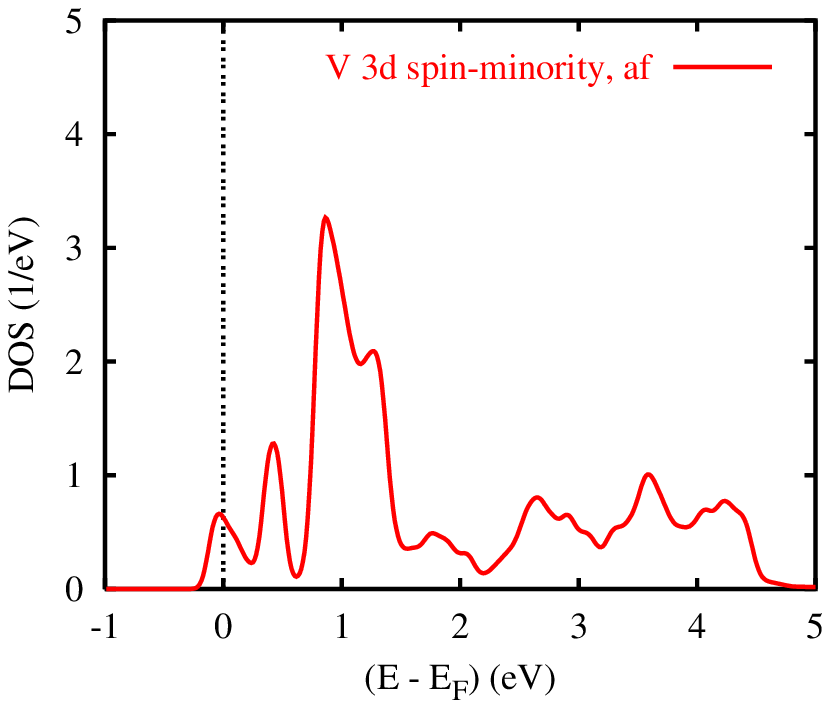}
\caption{(Color online) Partial V $3d$ spin-majority (top) and
spin-minority (bottom) DOS (per atom) for antiferromagnetic
intrachain coupling.}
\label{fig4}
\end{figure}

We next study the effects of the intrachain magnetic coupling on the
electronic structure of ferromagnetically coupled VO-chains. Partial
V $3d$ spin-majority and spin-minority densities of states for ferromagnetic
and antiferromagnetic exchange along the chains are shown in figures
\ref{fig3} and \ref{fig4}. The width of the spin-majority bands
in figure \ref{fig3} amounts to 4.8\,eV. It therefore is about
0.4\,eV larger than for the spin-degenerate bands, see figure \ref{fig2},
whereas the width of the spin-minority bands hardly alters.
In contrast, the spin-majority and spin-minority band widths are rather
similar in figure \ref{fig4}. While for ferromagnetic intrachain
coupling the spin-minority group of states is observed at higher
energies, leaving only V $3d$ spin-majority states occupied, both
spin components contribute at the Fermi level for antiferromagnetic
coupling. As a consequence, the local V magnetic moment accumulates
to only 0.71\,$\mu{\rm_B}$ in the latter case. Oxygen magnetic moments
are neglectible. On the contrary, the ferromagnetic coupling results
in magnetic moments of 0.92\,$\mu{\rm_B}$ for the V and 0.05\,$\mu{\rm_B}$
for the intrachain O sites, which sum up to 4\,$\mu{\rm_B}$ per unit
cell. Due to a strong spin splitting of nearly 0.8\,eV, see figure
\ref{fig3}, a large number of occupied states is shifted to lower
energies, paving the way for the ferromagnetic ground state.

In conclusion, electronic structure calculations using density
functional theory indicate that the experimental room-temperature
crystal structure of the zigzag spin-chain compound In$_2$VO$_5$ is incompatible
with both antiferromagnetic intrachain and interchain coupling.
Ferromagnetism is stabilized instead, which contradicts the crystal
chemical estimates by Volkova \cite{volkova07}. This discrepancy
probably traces back to hybridization between the V $3d$ and O $2p$
states, as reflected by remarkable oxygen magnetic moments. Nevertheless,
the antiferromagnetic coupling likewise comes along with energy gain
as compared to the non-magnetic solution. Because of narrow V $3d$ bands at the
Fermi energy, see figure \ref{fig2}, electronic correlations beyond
the local density approximation can play a role and further stabilize
antiferromagnetic interaction. However, this seems not to
be the case here, as recent experiments point at a transition from ferromagnetic
to antiferromagnetic exchange at low temperature, accompanied by structural alterations 
\cite{kataev07}. Since strong coupling of
the electronic system to the crystal lattice is typical for transition
metal oxides \cite{eyert05}, slight changes in the crystal structure
may induce relevant modifications of the magnetic exchange.
A large variety of phase transitions is known for compounds with
octahedrally coordinated transition metal atoms. The vanadium and
titanium Magn\'eli phases, for example, are subject to metal-insulator
transitions accompanied by distinct structural alterations
\cite{us03,leonov06}. A low temperature structural phase transition
in In$_2$VO$_5$ therefore still could cause an antiferromagnetic
ground state with frustrated zigzag spin-chains. In order to solve
this question, a detailed investigation of the In$_2$VO$_5$ crystal structure
is required.

\subsection*{Acknowledgement}
Valuable discussions with L.M.\ Volkova are
gratefully acknowledged. This work was supported by the Deutsche
Forschungsgemeinschaft (SFB 484).

\end{document}